# Design, fabrication, and performance of a versatile graphene epitaxy system for the growth of epitaxial graphene on SiC

S. Mondal,[1,2] U. J. Jayalekshmi,[1,2] S. Singh,[1] R. K. Mukherjee,[1,2] and A. K Shukla[1,2 a]

[1]CSIR-National Physical Laboratory, Dr. K.S Krishnan Marg, New Delhi-110012, India

[2]Academy of Scientific and Innovative Research (AcSIR), Ghaziabad-201002, India

**[a]** **Author to whom correspondence should be addressed:** shuklaak@nplindia.org

## Abstract

A versatile *Graph*ene *E*pitaxy (GrapE) furnace has been designed and fabricated for the growth of epitaxial graphene (EG) on silicon carbide (SiC) under diverse growth environments ranging from high vacuum to atmospheric argon pressure. Radio-frequency (RF) induction enables heating capabilities up to 2000°C, with controlled heating ramp rates achievable up to 200°C/s. Details of critical design aspects and temperature characteristics of the GrapE system are discussed. The GrapE system, being automated, has enabled the growth of high-quality EG monolayers and turbostratic EG on SiC using diverse methodologies such as close confinement sublimation (CCS), open configuration, polymer-assisted CCS, and rapid thermal annealing. This showcases the versatility of the GrapE system in EG growth. Comprehensive characterizations involving atomic force microscopy, Raman spectroscopy, and low-energy electron diffraction techniques were employed to validate the quality of the produced EG.

# I. Introduction

The experimental realization of graphene, a two-dimensional planar sheet of carbon atoms arranged in a hexagonal lattice, sparked widespread interest in the scientific community.[1] This interest stemmed from its extraordinary electrical, mechanical, thermal, and magneto-transport properties,[2–5] coupled with its versatile range of potential applications.[5–8] Despite exhibiting fascinating physical phenomena and high carrier mobilities, the challenges in reproducibly obtaining high-quality large-area domains hinder the practical use of exfoliated graphene obtained through micromechanical exfoliation of graphite.[9,10] Chemical vapor deposition (CVD) emerged as a frequent and cost-effective method for large-scale graphene production on metal foils. However, its limitations in achieving high crystallinity over large areas and the subsequent need for transfer to other substrates for device applications presented drawbacks.[7,11] The advent of epitaxial graphene (EG) grown on SiC offered a solution by addressing these issues. EG is formed through Si sublimation from the first few SiC layers at high temperatures in ultra-high vacuum (UHV) and the consequent rearrangement of the remaining carbon atoms into a honeycomb lattice.[12–15] EG is single crystalline, homogeneous over a large area, and does not need any transfer for device fabrications as it is already anchored on the large band gap SiC.[12–15] Sublimation growth of EG on SiC at high temperatures (typically ≥ 1200°C) in UHV environment paved the way to realize unique electronic band structure of graphene.[16,17] However, the suitability of UHV-grown EG for specific uses, such as the quantum Hall resistance standard (QHRS), the primary resistance standard based on the quantum Hall effect, was compromised due to intrinsic structural defects.[12] Moreover, its restricted uniformity, specifically in achieving a consistent monolayer coverage over a considerable area alongside bilayer patches, stemmed from the rapid escape of Si atoms, leading to a non-equilibrium growth scenario.[18,19] A notably high silicon-to-carbon ratio exists at lower temperatures commonly employed for EG growth in ultra-high vacuum (UHV) environment. This high Si/C ratio is due to the substantial disparity in vapor pressures between silicon and carbon, leading to the formation of smaller graphene domains. This occurrence happens as there is limited time for carbon atoms to diffuse over the SiC surface. However, the ratio of silicon to carbon-containing species decreases as the temperature rises, fostering a more uniform and controlled growth of EG under near-equilibrium conditions.[15,20–22] Consequently, three distinct research groups reported the production of epitaxial graphene (EG) by thermally sublimating silicon carbide (SiC) at significantly higher temperatures. Their findings displayed notable enhancements in morphology and thickness uniformity over a large scale. For instance, Virojanadara et al.[23] and Emtsev et al.[18] conducted EG growth on SiC at 2000°C and 1650°C, respectively, under ~1 atm pressure of Ar gas by using radio frequency (RF) induction-heated commercial cold-wall SiC growth reactors. These reactors featured specially designed graphene susceptors where the SiC chip was placed, and for the remainder of this report, we refer to this growth configuration as the "open configuration." The rationale behind the improved EG growth in this open configuration was attributed to the combination of high growth temperatures and the presence of atmospheric pressure of Ar gas, which effectively suppressed the escape of silicon vapor.[18,23] Later on, an automated inductively heated hot wall reactor was reported for open configuration EG growth at 1650°C in ~1 atm Ar gas ambient.[24] Heer et al. introduced a different technique, known as confinement-controlled sublimation (CCS), to minimize the escape rate of silicon vapor during sublimation for epitaxial graphene (EG) growth. This method involved using a specially designed graphene crucible with a calibrated leak. High-quality EG monolayers and multilayers were grown on SiC at growth temperatures exceeding 1500°C using the CCS method, where EG growth rates could be adjusted over $\sim 10^6$ times compared to typical UHV growth with calibrated leak providing $\sim 10^3$ times suppression of EG growth rate and additional $\sim 10^3$ times control on growth rate suppression could be achieved by introducing atmospheric pressures of Ar gas.[19] Subsequently, various research groups explored diverse growth approaches in both open and CCS configurations to enhance the morphology, reduce defects, achieve better homogeneity, and refine the electronic properties of monolayer EG.[25–28] Expensive equipment such as modified cold wall commercial SiC reactors and graphite-lined vacuum furnaces have been predominantly utilized,[18,23,26] followed by the development of relatively smaller and more cost-effective hot wall reactors for EG growth in open configuration.[24,29] Conversely, CCS growth of EG has mainly occurred in non-commercial home-built RF induction heating-based growth systems.[19,30–33] Comprehensive design and construction details of these growth systems, pertaining to both open and CCS configuration growth of EG, have either not been adequately provided or are fragmented across various reports and doctoral theses. Additionally, detailed thermal characteristics of these growth systems are scarcely available.

Silicon carbide (SiC) is a polar material presenting two distinct faces: Si-face (0001) and C-face (000$\bar{1}$) each resulting in different types of graphene growth. On the Si-face, graphene grows much slower than on the C-face, enabling better control over thickness and homogeneity. Monolayer EG grows on the Si-face with Bernal stacking (AB staked), and it is very useful for QHRS and various applications.[34–36] However, it exhibits limited carrier mobility due to an interface buffer layer, making it less appealing for many electronic device applications.

In contrast, EG grows on the C-face without an interface buffer layer. It tends to grow in a multilayer form with random rotational orientation between individual layers stacked in a non-Bernal fashion. The uppermost graphene layer typically exhibits monolayer graphene electronic behavior.[37] This type of multilayer EG, termed turbostratic graphene, possesses significantly higher carrier mobilities than Bernal-stacked EG.[38,39] The disorder and misalignment between layers can alter the dispersion of electronic states and open the bandgap.[40,41] The attractive electronic properties of turbostratic graphene make it superior for various applications such as energy storage devices, superconductors, sensors, and electronic devices.[42,43] Additionally, the stacking structure of the turbostratic multilayer graphene can decrease the effect of attachment of charge impurities and surface roughness.[44]

Rapid Thermal Annealing (RTA) has emerged as an important technique for turbostratic graphene synthesis, offering shorter growth times due to high annealing rates. It reduces charge transfer between graphene and SiC by weakening their interaction, significantly reducing sheet resistance and step bunching in turbostratic EG.[45] Various heating sources and strategies such as halogen lamps,[46] pulsed laser,[45] high-temperature flash annealing,[47] CVD,[48] arc discharge,[49] laser-assisted CVD,[50] microwave plasma enhanced CVD[51] and infrared lamp heating based commercial system[52] have been employed for RTA-based synthesis of turbostratic graphene. However, many of these RTA systems (HT flash annealing, CVD, Halogen lamp) have limitations in terms of achieving high annealing rates (>100°C/s) at high temperatures (≥1800°C) in a wide range of growth environments (HV to atmospheric pressures of Ar gas). Even though pulsed/continuous laser-based RTA systems can achieve very high annealing rates, these have issues with choosing a particular laser wavelength and laser power to avoid the possibility of burning SiC and controlling heating rates.

While RF induction provides for fast, localized, and controlled heating, there are hardly any reports on RF induction heating-based RTA systems capable of growing turbostratic EG at high growth temperatures and heating rates.

In this article, we report the design, construction, and performance of an RF induction heating-based system for *grap*hene *e*pitaxy (GrapE), a versatile system tailored for EG growth on SiC. While building upon earlier pioneering works,[18,19,23] our goal is to provide a comprehensive insight into the design, construction, and capabilities of a system adaptable to various EG growth methods. The GrapE furnace was designed to meet the primary requirements for EG growth on SiC: (i) offering controlled growth environments from high vacuum (HV) to atmospheric pressures using chosen gases (Ar/Ar+H2); (ii) enabling growth temperatures up to ~2000°C; and (iii) allowing closed-loop control of heating rates from 1°C/s to 200°C/s. Its versatility lies in implementing different popular EG growth techniques within the same system, such as open configuration, CCS, and RTA. We offer comprehensive insights into the design and construction of the GrapE system, focusing on essential design prerequisites. Detailed discussions on the thermal characteristics of GrapE are also provided. The growth furnace is automated to ensure high-quality EG production on SiC. Additionally, we showcase a modified CCS growth approach—polymer-assisted CCS—to achieve shallow step heights (< 1 nm) on SiC(0001). Furthermore, we employ RTA with a remarkably high annealing rate (200°C) to produce turbostratic EG on SiC($000\bar{1}$) and analyze the effect of the growth environment on its stacking order. This work introduces the novel use of such high RTA heating rates to grow turbostratic EG in a typical RF induction heated hot wall reactor-like system, which, to our knowledge, has not been previously explored. The quality of the produced EG is established through comprehensive characterizations utilizing atomic force microscopy (AFM), Raman spectroscopy, and low-energy electron diffraction (LEED).

## II. Design and construction

### A. General construction and vacuum handling part

A 3D isometric view of the GrapE induction furnace, generated using SolidWorks 2020, is shown in Fig. 1. A modular frame made of aluminum extrusion profiles has been designed to support the GrapE system. The GrapE furnace has been constructed around a stainless steel (SS) DN 40 CF six-way cross piece (A) fixed to the system frame via a custom-built aluminum support. A dome-shaped quartz tube (B) serves as the reaction chamber and houses a graphite crucible assembly (C). This quartz dome is connected to one of the four horizontal ports of a six-way cross through a specially designed DN 40 CF-Wilson seal port (D) supplied by Excel Instruments, India. Design and actual pictures of the quartz dome assembly comprising Wilson seal,[53] DN40CF adapter, and quartz dome are shown in Fig. S1(Supplementary material). Use of Wilson seal allowed us to open/re-mount and change the quartz dome easily and quickly. Two strategically positioned high-speed DC fans (E; SERVO-G123B24BBZIP-00, RJ, USA) provide air-cooling to the quartz dome.

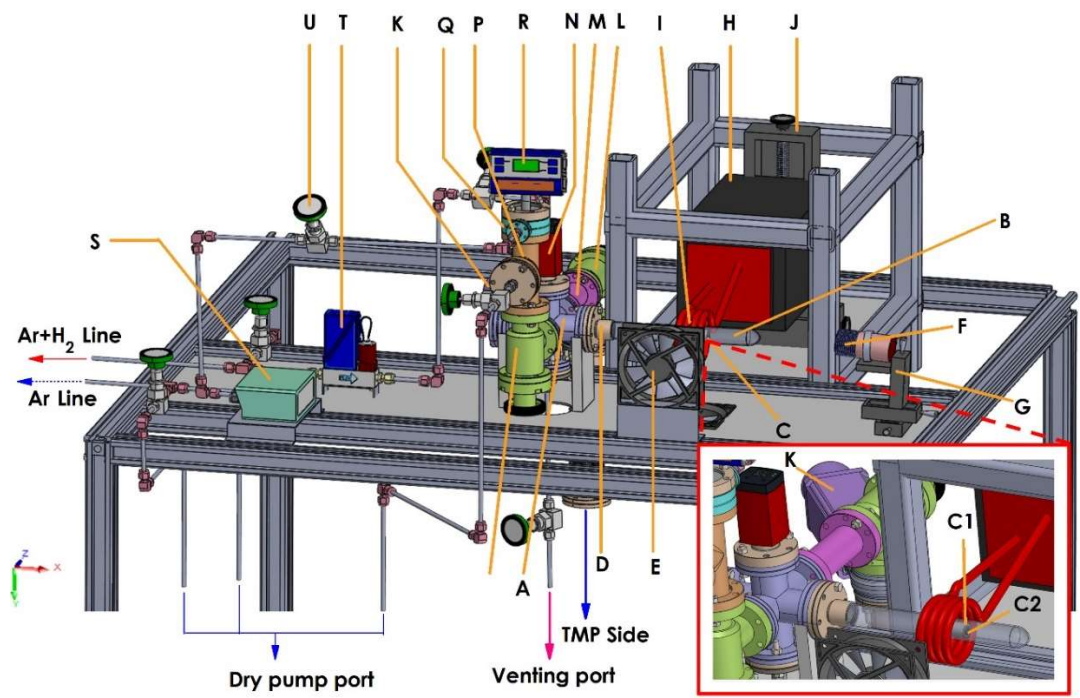

*Figure 1:* *3D isometric view of GrapE system. (A) DN40CF six-way cross, (B) quartz dome reaction chamber, (C) graphite crucible surrounded by graphite felt insulation, (D)Wilson seal, (E) high-speed DC cooling fan, optical pyrometer, (F) optical pyrometer, (G) XYZ linear translational mounting stage for pyrometer, (H) RF remote heat station, (I) RF induction coil, (J) XYZ linear translational mounting stage for RF remote heat station,(K) pneumatic gate valve (also shown in the inset), (L) HV right angle isolation valve, (M) DN 40 CF straight nozzle, (N) cold cathode gauge, (O) HV right angle isolation valve, (P) Tee, (Q) special double sided/spacer DN 40 CF flange with two DN 16 CF ports, (R) convection gauge,(S) Controlled gas delivery system,(T)mass flow controller(MFC),(U) needle valve. Part of the gas lines attached to a dry pump, the chamber pumping port connected to a turbomolecular pump (TMP), and the venting port of the chamber attached to a needle valve are also shown. The inset shows a magnified view of the quartz dome growth chamber, housing graphite crucible, assembly: (C1) graphite rigid felt insulation and (C2) graphite crucible.*

Such air-cooling was necessary to maintain the structural stability of the quartz dome during high-temperature operations, mainly when the furnace operates at temperatures ≥ 1600°C for a longer duration (≥ 30 min). An infrared optical pyrometer (F; CTM-1SF75H1, Micro-Epsilon, Germany) with a temperature range of 800 – 2200°C is employed for the online graphite crucible temperature measurement. The pyrometer is equipped with two laser beams to accurately locate and fix the temperature measurement point; the typical spot size is ~ 1mm at our working distance of ~ 16cm. The pyrometer is mounted on an *XYZ* stage (G) to position the temperature measurement spot on the graphite crucible precisely. *XY* and *Z* travel ranges of this translational stage are 50 and 25 mm, respectively, with a straight-line accuracy of 10 μm. A 6-kW radio frequency (RF) induction heating system comprising an RF power supply and remote heat station (H; 5060 LI, Ambrell, USA) has been used to facilitate the heating operations of the GrapE furnace. A water-cooled copper RF induction coil (I), connected to an RF remote heat station, is used for RF induction heating of graphite crucible. The remote heat station is mounted on top of another *XYZ* stage (J) so that RF coil can be precisely and safely translated in/out over the quartz dome area housing graphite crucible assembly. *X, Y* and *Z* travel ranges of this translational stage are 150 mm with a straight-line accuracy of 20 μm. Both the *XYZ* translational stages (G and J), manufactured by Scientific Components, India, are anchored at the system frame. A pumping and gas delivery system has been designed to facilitate the operation of the GrapE system in varied environments ranging from high vacuum (HV) to atmospheric pressures of Ar/Ar+$H_2$ gases. The pumping system of the GrapE furnace is based on a turbomolecular pump (TMP; Hi-Pace 300, Pfeiffer Vacuum, Germany) backed by a double-stage rotary vane (DUO 11M, Pfeiffer Vacuum, Germany) roughing vacuum pump. A DN 40 CF pneumatic ultra-high vacuum (UHV) mini gate valve (K; series 01.0, VAT, Switzerland) is connected to TMP through a series of UHV components (conical SS DN 100-DN 63 CF reducer, zero length DN 63-DN40 CF reducer, and DN 40 CF bellow) not shown in Fig. 1. This pneumatic gate valve is attached to one of the ports of an HV right angle valve (L; 951-5091, Varian Inc. Vacuum Technologies, USA) which in turn is connected to one of the horizontal ports of six-way cross via a straight DN 40 CF nozzle (M). This hand-operated right-angle valve allows us to slowly connect growth chamber of the GrapE furnace to TMP, running at full speed after evacuation by dry roughing pump (not shown in Fig. 1), while the pneumatic gate valve provides immediate isolation/connection of TMP pumping from the growth chamber. A cold cathode gauge (N), (PKR 361 from Pfeiffer Vacuum, Germany) has been mounted on the top port of the six-way cross via a curved DN 40 CF elbow to measure the pressure of the GrapE system. The measuring range of this cold cathode gauge is $1 \times 10^{-9}$ to $1 \times 10^{3}$ mbar. We routinely get $\leq 5 \times 10^{-7}$ mbar base pressure of the GrapE system within a few hours (~ 4 – 6 h) of TMP pumping, and base pressure can be improved to $\leq 8 \times 10^{-8}$ mbar after ~ 24 h pumping.

We have also achieved much lower ($\leq 5 \times 10^{-9}$ mbar) base pressure after a bakeout (~120°C, 24 h), showing that the GrapE system can have a UHV environment as well if needed. In the case of bakeout, we replaced DN 40 CF-Wilson seal-quartz dome assembly with a UHV-compatible quartz dome welded to an SS DN 40 CF flange (GMQS100F3RN from Kurt J. Lesker, USA). However, most EG growth runs only need $\leq 5 \times 10\text{-}7$ mbar base pressure. Therefore, the required time for loading a new sample and starting EG growth in the GrapE furnace is just ~4 - 6 hours. The bottom flange of the six-way cross is dedicated to controlled venting of the GrapE furnace with Ar gas through an SS bellows sealed needle valve (SS-4BG from Swagelok, USA). A magnified view of the quartz dome growth chamber housing all graphitic crucible assembly (C) is shown in the inset of Fig. 1. Isostatic, semiconductor grade, high-density graphite has been used to fabricate specially designed graphite crucible (C2), and it has been placed inside a rigid graphite felt cylinder (C1). We have used two different sources of graphite rods (T-6, Ibiden, Japan, and HPD, Entegris, USA) to machine our graphite crucibles having equivalent performance during EG growth. The low density, low thermal conductivity, and porous nature of rigid graphite felt (FU 2914, Schunk GmbH, Germany) make it ideal for thermal insulation of graphite crucibles and minimize radiation heat loss, especially at high-temperature operations.

## B. Gas handling part

The GrapE growth chamber is connected to the gas injection part through another hand-operated right-angle HV isolation valve (O) attached to one of the horizontal ports of the six-way cross. A DN 40 CF Tee-piece (P) is connected to the vertical port of the right-angle valve (O). A special double-sided/spacer DN 40 CF flange with two DN 16 CF ports (Q; 420FDP040-2-16, Pfeiffer Vacuum, Germany) is mounted on top port of Tee-piece (P). This double-sided flange's two DN 16 CF ports serve as separate gas inlet ports for Ar and Ar+$H_2$ gases. A convection gauge with integrated controller and display (R; KJL300 series, Kurt J. Lesker, USA) is attached to the top side of the double-sided flange for the monitoring of pressure during pumping, purging, and filling of the gas line portion of the GrapE furnace. The measuring range of this convection gauge is $10^{-4}$ to 1,333 mbar, allowing us to monitor atmospheric pressure fillings of gases in the GrapE furnace. A multi-stage dry root pump (ACP 15 from Pfeiffer Vacuum, USA), not shown in Fig. 1, is connected to the horizontal port of the Tee-piece through a Swagelok needle valve to facilitate pumping and purging of the gas handling part as well as to evacuate GrapE furnace to rough vacuum, whenever needed.

One of the DN 16 CF gas inlet ports is connected to the output channel of a specially designed controlled gas delivery system (CGDS) (S) through a Swagelok needle valve(U). The CGDS system, supplied by MCQ srl, Italy, can maintain controlled gas pressures in the range of 0.5 to 3 bar with varying gas flow rates. A needle valve connects another DN 16 CF gas inlet port to a mass-flow controller (T) output. Two separate Swagelok Tee pieces are connected to the input channel of CGDS and MFC. Two other ends of these tee pieces are connected to a dry vacuum pump and respective gas cylinders (not shown here). Input channels of CGDS and MFC are connected to 6N purity Ar gas cylinder and 5.5N purity Ar(95%) +$H_2$(5%) gas mixture cylinder, respectively. Two separate needle valves, one between the dry pump and input channel of CGDS/MFC and another between the respective gas cylinder regulator and CGDS/MFC, are also installed to facilitate controlled pumping/purging of these lines. Our modular gas handling design helps us to pump/purge all the gas lines efficiently, even up to the mouth of gas cylinders. All the SS (316/304) components (1/4" seamless tubing, tee-pieces, unions, elbows, sealing ferrules, and needle valves) used in gas line fabrication have been sourced from Swagelok, USA. All metal, leak-free, bakeable gas lines can attain low base vacuum ($\leq 5 \times 10^{-5}/\leq 1 \times 10^{-7}$ mbar without/with baking) using combination pumping of dry and TMP pump. Such good base pressure of the gas handling part is beneficial to maintain the purity of injected gases in the GrapE furnace. Figure 2 exhibits the actual photograph of the GrapE system, and all the components mentioned in the discussion of Fig. 1 are marked here. Figure S2 and S3(a) (supplementary material) contain more pictures of pumping and gas line arrangements. Additional 3D isometric CAD views of different sides: An Actual photograph of a close-up view of the quartz dome growth chamber, housing graphite crucible assembly, and induction coil placed around it is given in Fig. S3(b) (supplementary material).

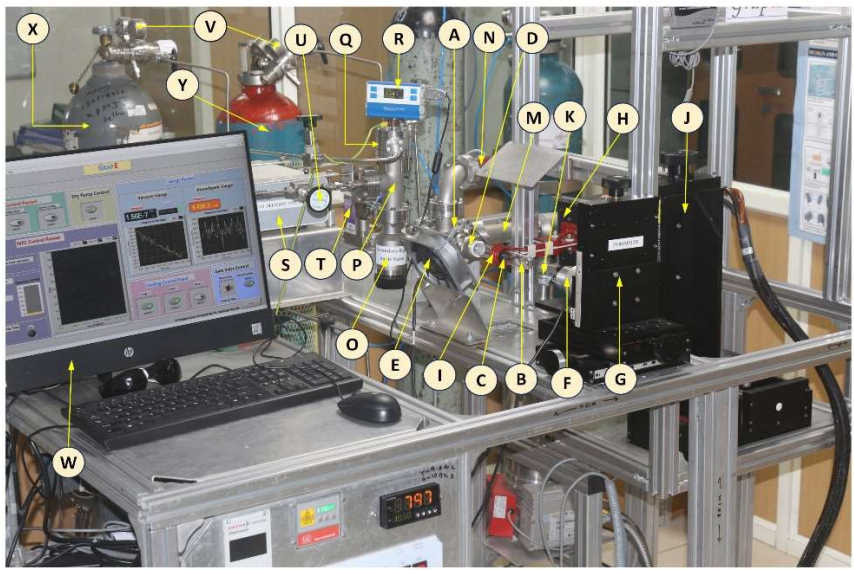

*Figure 2*: Actual photograph of GrapE growth system. (A) DN40CF six-way cross, (B) quartz dome reaction chamber, (C) graphite crucible surrounded by graphite felt insulation, (D) Wilson seal, (E) high-speed DC cooling fan, optical pyrometer, (F) optical pyrometer, (G) XYZ linear translational mounting stage for pyrometer, (H) RF remote heat station, (I) RF induction coil, (J) XYZ linear translational mounting stage for RF remote heat station,(K) pneumatic gate valve, (M) DN 40 CF straight nozzle, (N) cold cathode gauge, (O) HV right angle isolation valve, (P) Tee, (Q) special double sided/spacer DN 40 CF flange with two DN 16 CF ports, (R) convection gauge, (S) Controlled gas delivery system, (T) Mass flow controller, (U) needle valve, (V) 6N Gas cylinder regulator,(W) PC for growth parameter measurements and automatic control of GrapE (X) 6N Argon cylinder (Y) 5.5N Ar+$H_2$ cylinder.

## C. Graphite crucible/susceptor

One of the most critical components of the GrapE system is the graphite crucible/susceptor housing SiC chip on which EG is grown. It acts as an enclosure/susceptor for SiC during the thermal sublimation of Si out of SiC and consequent EG growth. We have employed two types of growth strategies: growth in CCS and open configurations. For CCS growth, the design and dimensions of the graphite crucible are crucial in controlling EG growth rate and quality.[19,30–32] Although several reports of EG growth using the CCS method have come out since the introduction of this very popular growth method, details of the exact design and dimensions are scattered.

Considering that crucible design is paramount for CCS growth method, we provide detailed CAD drawings and a 3D view of this crucible in Fig. 3. Graphite crucible is comprised of two parts, a lid, and one side open cylinder, and it facilitates Si sublimation-induced EG growth on SiC in confined geometry. A cross-sectional 2D CAD drawing of the lid is shown in Fig. 3(a). The lid has been designed to tightly close the graphite cylinder's open end. 1 mm diameter through hole has been drilled at the center of the lid to provide a controlled leak of Si vapor from the crucible during thermal sublimation induced EG growth on SiC. Figure 3(b) exhibits a cross-sectional 2D CAD drawing of the graphite cylinder. A blind hole (2.4 mm diameter; 3 mm depth) has been drilled at the bottom center of the graphite cylinder to position the pyrometer spot, enabling temperature measurements closely resembling the temperature inside the crucible. Figure 3(c) and (d) show 3D isometric views of the graphite lid and crucible, respectively, and Fig. 3(e) shows an assembled view of the graphite crucible. The modular design of the crucible allows lids with different diameter leak holes on the same cylinder to tune the Si escape rate and consequent EG growth.[19,30–32]

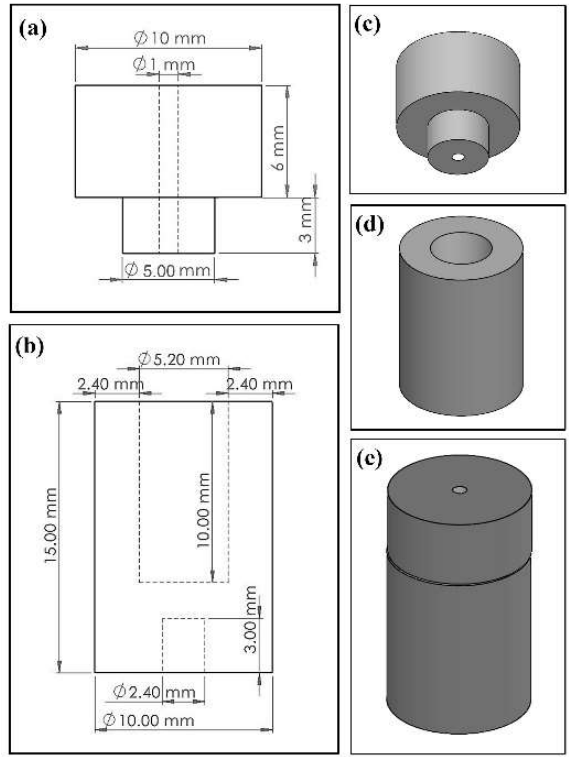

*Figure 3:* CAD design and 3D view of graphite crucible for CCS growth configuration. (a) Cross-sectional 2D drawing of the lid of the crucible. Dotted lines have marked a leak hole of 1 mm in diameter. (b) Cross-sectional 2D drawing of the lower cylindrical part of the crucible. The inner cavity (length: 10 mm, diameter: 5.2 mm) has been shown by dotted lines. A blind hole (2.4 mm diameter) is also marked at the bottom of the cylinder by dotted lines. (c) 3D isometric view of the lid. (d) 3D isometric view of cylinder. (e) 3D isometric view of crucible assembly (lid + cylinder). All dimensions are given in mm.

EG growth in open configuration does not depend on crucible geometry or Si leak controlling parameters because the susceptor is open from both sides in this process. CAD drawings of the graphite susceptor, designed for open configuration growth, are shown in Fig. 4. It has two parts: a cylinder and a sample carrier. Figure 4(a) shows a 2D cross-sectional drawing of both sides of an open cylinder with a slot running along its entire length, as depicted in a 3D isometric view in Fig. 4(c). Figure 4(b) exhibits the 2D top view of the tray-shaped sample carrier, and the sample placement area is marked therein by a shaded region. Its 3D isometric view is shown in the Fig. 4(d). The sample carrier tray has a 2 mm deep depression to safely place the SiC chip in standard standalone geometry or face-to-face/face-to-graphite configurations (Fig. 4(b)). Open-ended cylinder and sample carrier have been designed so that the sample carrier snuggly fits in the cylinder slots, as shown in the assembled view of the susceptor in Fig. 4(e). All the graphite crucibles/susceptors reported here are machined at the mechanical workshop of CSIR-NPL.

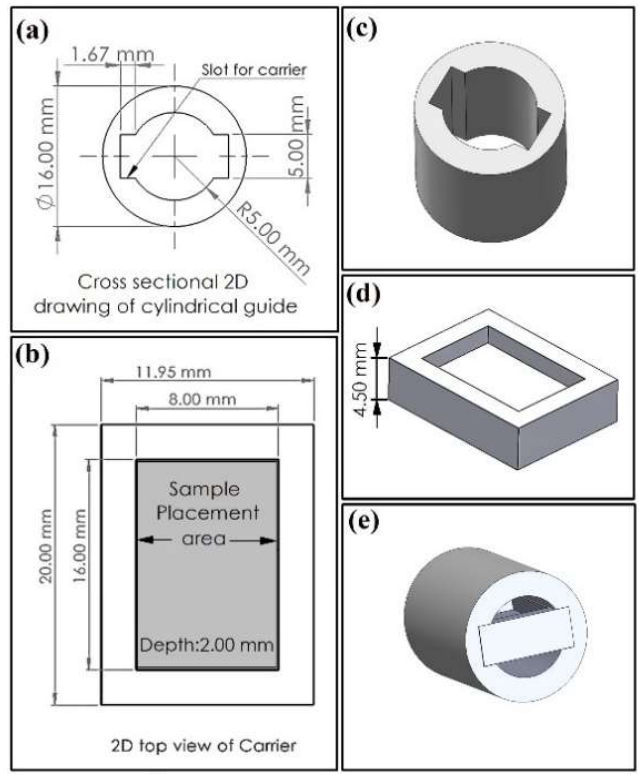

*Figure 4:* *CAD design of graphite susceptor for Open growth configuration. (a) cross-sectional 2D drawing of the cylindrical guide. (b) Top view of 2D drawing of the sample carrier. The shaded region marks the sample placement area. (c) 3D isometric view of the cylindrical guide of sample carrier. (d) 3D isometric view of the sample carrier. (e) 3D isometric view of open crucible assembly (cylindrical guide + sample carrier). All dimensions are given in mm.*

## III. Operation and Thermal characteristics:

The efficacy of any EG growth system heavily relies on its capability to attain stable target temperatures within specified environments, precisely control temperature ramp rates (both upward and downward), and ensure uniform temperature dispersion within the graphite crucible.

As mentioned earlier, the GrapE furnace can operate in a wide range of environments, from high vacuum to atmospheric pressures. In this section, we delve into the automated control mechanism and the thermal attributes of this growth system.

**A. Automatic control and measurement**

LabVIEW 2019 (NI, USA) has been used to implement automation/interfacing of different components of the GrapE system. Figure 5 shows the schematic of the LabVIEW-based automatic control and growth parameter (process temperature, pressure, and gas flow) measurement interface. All the vacuum pumps, gauges, isolation pneumatic gate valve, CGDS/MFC for gas pressure/flow control, cooling fans, and temperature measurement/control components are connected to GrapE's PC through a multifunctional NI DAQ card (NI-USB-6343).

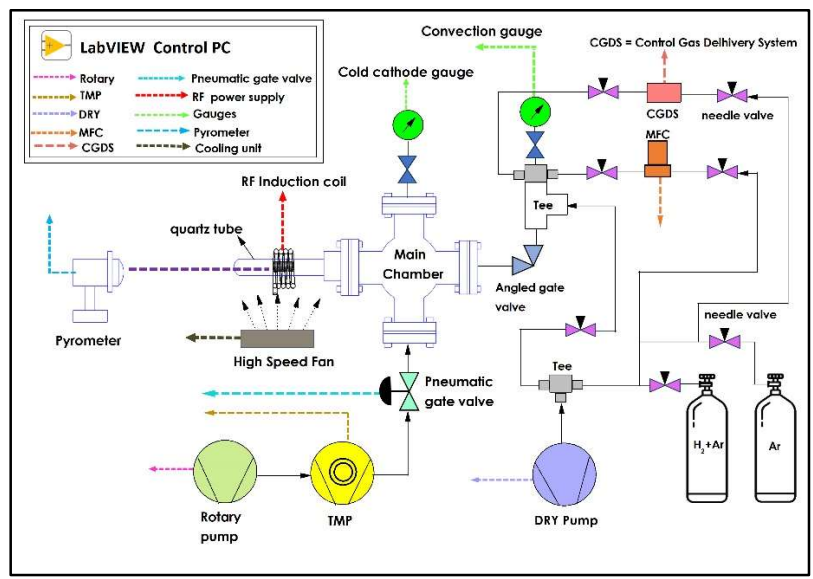

*Figure 5:* Schematic of NI-LabView-based automatic control and temperature, pressure, and gas flow measurement in the GrapE furnace.

Closed-loop temperature control is implemented to achieve and maintain a stable growth temperature by connecting a proportional-derivative-integral (PID) temperature indicator/controller to the RF power supply and optical pyrometer. The optical pyrometer measures process temperature. This information is fed to the PID controller, which tunes the power of the induction heating coil based on a dynamic response from a feedback control circuit.

**B. Thermal performance:**

All the thermal characteristics have been tested using a CCS Graphite crucible assembly placed inside the quartz dome growth chamber. A water-cooled RF induction heating coil has been placed around the quartz dome to cover more than the entire length of the graphite crucible and provide uniform heating. RF induction heating provides localized, fast, non-contact, and clean heating. The measuring spot of the pyrometer has been precisely positioned inside the blind hole drilled at the bottom center of the graphite crucible to have accurate temperature measurements. A stable temperature with a narrow distribution along the length and diameter of the graphite crucible over the process duration is an essential factor for the sublimation growth of EG.[15,54] Design and heating arrangement of our system is inspired by typical hot wall SiC growth reactors. It is challenging to know the exact temperature distribution in such systems because of very compact and closed rector designs, and simulated models are mainly employed to understand temperature distribution on the graphite crucible.[15,54,55] To the best of our knowledge, beyond such simulations, we have not come across any reported experimental determination of temperature distribution on the crucible used for EG growth. We have performed temperature measurements at a fixed RF power on several points of the graphite crucible bottom surface to understand the actual temperature distribution along the diameter of the growth crucible. Figure 6(a) represents the methodology of collecting data at *XZ* planes across the bottom surface of the graphite crucible over a diameter of ~10mm. Eleven grid lines 1 mm apart from each other along the X and Z axes have been created to record the *XZ* plane temperature data. All the temperatures measured at various points of the *XZ* plane have been converted into polar coordinates $(r, \theta)$ for better representation of temperature distribution. For example, the red star marker on Fig. 6(a) indicates a given *XZ* coordinate (X: 2mm, Z: 3.5mm) of the laser spot of the pyrometer at the bottom of the crucible. The polar coordinate of this particular *XZ* point corresponds to r = 4 mm and θ = 30°. An RF power has been fixed to achieve 1907°C (arbitrarily chosen) at the crucible's bottom center (r = 0 mm). The measured temperature contour plot in polar coordinates is given in Fig. 6(b), representing radial temperature distribution over the diameter of the crucible. The measured temperature variation is 6°C (~0.3%) across the total diameter of the crucible and only 3°C (~0.15%) across the inner cavity radial distance (diameter: 5.2 mm) of the crucible within which SiC chip is placed. The measured temperature distribution of our graphite growth crucible is comparable to the reported simulated temperature distribution[15,54] and even better than the experimentally measured temperature distribution[55] of the growth crucible.

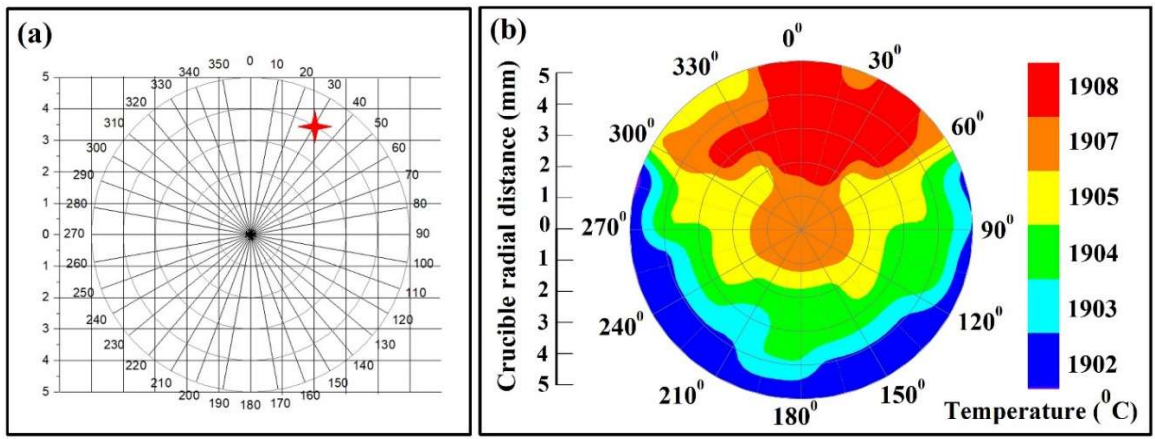

*Figure 6:* Radial temperature distribution map of graphite crucible. (a) schematic of the methodology for generating the data point to map the temperature in polar coordinate (r,θ) over the diameter of the graphite crucible (bottom). The star mark (red color) indicates the focused laser spot position (r: 4mm & θ: 30°) at the bottom of the crucible. (b) The experimentally determined temperature distribution of the graphite crucible from the center to the outer perimeter at different angles have been plotted in polar coordinates.

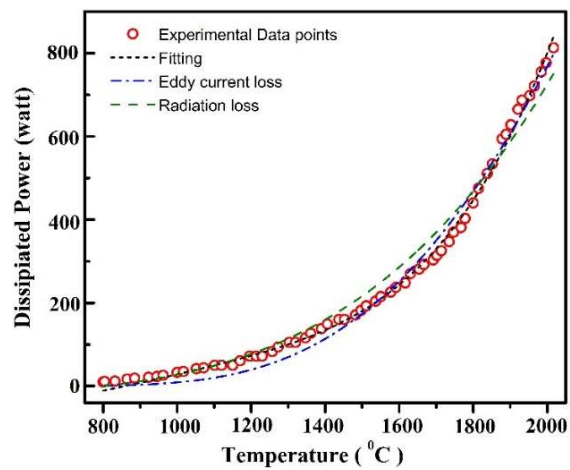

*Figure 7:* *Variation of dissipated RF power with temperature of the graphite crucible. Experimental data (open circles, red), fit to the data (solid line, black), eddy current loss (dot-dashed line, blue), and radiation loss (dashed line, green) are shown.*

Figure 7 shows the variation of dissipated RF power ($P_d$) applied to induction coil with stabilized or equilibrium temperature (T) of the graphite crucible. Variation of $P_d$ with T can be modeled by considering mainly radiation heating loss and eddy current loss, as expressed in Eq. 1.[56,57]

$$P_d = \frac{k}{\rho} exp\left(-\frac{Q}{k_B T}\right) + \sigma A(T^4 - T_0^4)$$

$$= a\, exp\left(-\frac{b}{T}\right) + c(T^4 - T_0^4) \tag{1}$$

Here, the first term is related to the eddy current loss of heating due to the induced eddy current generated on the surface of the graphite crucible placed inside the induction coil. $\rho$ is the electrical resistivity, Q is the activation energy, $k_B$ is the Boltzmann constant, and $k$ is the constant against $\rho$, depending on crucible geometry, magnetic flux density, and frequency. The second term represents radiation loss due to radiation heating from a black body like a graphite crucible. Here, $\sigma$, $A$, and $T_0$ are the Stephan-Boltzmann constant, the graphite cylinder's surface area, and the surroundings temperature, respectively. We have performed iterative fitting of $P_d$ (T) (Fig. 7) using Eq. 1 to understand the thermal characteristics of the graphite growth crucible. We can only record T ≥ 800°C due to the measurement limits of the used pyrometer, and therefore $T_0$ is set to 800°C. Proportionality constants a, c, and temperature-dependent activation energy linked parameter b (Eq. 1) have been freely varied to obtain the best fitting. It is evident from Fig. 7 that the radiation loss component (dashed line, green) models $P_d$ (T) variation well for T ≤ 1450°C but deviates from the trend for higher temperatures. On the other hand, the Eddy current loss component (dashed-dot line, blue) can reasonably fit the $P_d$ (T) variation for T ≥ 1450°C but shows deviation for lower temperatures. We get the best fitting of $P_d$ (T) variation by using a combination of both the components, radiation, and Eddy current loss (Fig. 7). For the best fit (solid line, black), we obtain $a = 66149.62$ W/$\Omega m$ and $c = 4.65 \times 10^{-11}$ W/°C$^4$. We calculate the electrical resistivity of graphite to be $15\mu\Omega m$ by using the value of $a$. Considering the dimension of the graphite crucible and value of $c$, we estimated the radiative loss constant to be $5.4 \times 10^{-8}$ W/ m$^2$ °C$^4$, very close to the Stephan-Boltzmann constant for a black body.[58] Fitted curve shown in Fig. 7 can be also utilized to determine the required RF power to reach a specific temperature on the graphite growth crucible. Figure S4 (supplementary material) shows the variation of applied RF power with measured temperature of graphite growth crucible with and without rigid graphite felt used for thermal radiation shielding. We can immediately realize the importance of placing the graphite crucible inside a rigid felt cylinder as the required RF power to attain a specific temperature decreases drastically when graphitic rigid felt thermal insulation is placed over the graphite growth crucible. For example, the required RF power to reach ~1960°C reduces from ~1.4 kW for the case of the bare crucible to ~0.7 kW for the crucible with thermal insulation. This significant reduction in RF power is due to the minimization of radiation loss when a rigid felt radiation shield is used, especially for temperatures ≥ 1450°C (Fig. S4, supplementary material). The identical heating and cooling rate trend indicates excellent control over temperature ramping up and down in the GrapE system. (Fig. S5, supplementary material).

## IV. Growth performance

We will now delve into the epitaxial graphene (EG) growth capacity of the GrapE system in various growth configurations, having first established its effectiveness in achieving stable high temperatures across different environments with controlled ramp rates. Semi-insulating 4H-SiC (Cree Inc., USA) has been utilized for all EG growth procedures in this study. Unless otherwise specified, the Si-face of the SiC has been the chosen surface for all growth processes. The process involved dicing 5×5 mm SiC chips from a 4" SiC wafer and protecting the Si-face using AZ1518 photoresist during dicing to prevent chipping. After dicing, thorough photoresist removal was ensured by repeated ultrasonication in boiling acetone/IPA, followed by additional ultrasonication in acetone/IPA at room temperature, before loading the SiC chips into the GrapE system for graphitization. The detailed recipe for EG growth in open, CCS, and polymer-assisted CCS (PACCS) can be found in Fig. S6 (supplementary material). In brief, the EG growth process comprises four steps: (i) SiC surface cleaning involving annealing in vacuum up to 1050°C to eliminate adsorbates/oxides, followed by (ii) gas (Ar) insertion at 1 atm pressure and annealing at 1050°C for further cleaning; (iii) Graphitization under 1 atm Ar at 1450°C to foster a uniform buffer layer on SiC, and (iv) graphene growth at 1850°C.

### A. Epitaxial graphene growth in open configuration:

Pre-cleaned 4H-SiC chip is placed at the center of the rectangular graphite carrier of graphite susceptor (Fig. 4) along with a few dummy SiC chips surrounding it, and graphene growth is carried out by following the growth profile shown in Fig. S6 (supplementary material). AFM is performed in tapping mode using the Bruker Multimode 8 system. Figure 8(a) exhibits the AFM topographic image of grown EG, and it reveals an irregular step-terrace like morphology similar to earlier reports of open configuration growth.[18,26] Terrace width varies between 0.4 - 0.5 μm and step height varies from 0.5 - 0.6 nm as evident from the height profile along AB line given in the inset. Observed step heights are close to half of the unit cell height (1 nm) of 4H-SiC. It is interesting to note that step heights observed in our case are much smaller than typical step heights (~6 – 15 nm) observed for EG grown in open configuration.[18,26] We can also observe some randomly distributed pit-like morphological defects. The average root mean square (*rms*) surface roughness within individual terraces is ~0.04 nm. Raman spectroscopy measurements have been performed using Renishaw inVia Raman microscope with 50X objective (0.5 NA), 2400 grooves/mm diffraction grating, and 532 nm excitation wavelength. Figure 8(b) depicts the Raman spectrum, and characteristic G and 2D peaks are located at ~ 1602 and ~ 2711 cm$^{-1}$, respectively. Weak intensity, broad defect-related D, and buffer layer features are distributed between 1300 and 1460 cm$^{-1}$. The D peak, usually observed at ~1350 cm$^{-1}$, is almost non-existent or very small due to very little disorder or defects on the graphene sheet. Raman features of SiC, distributed over a

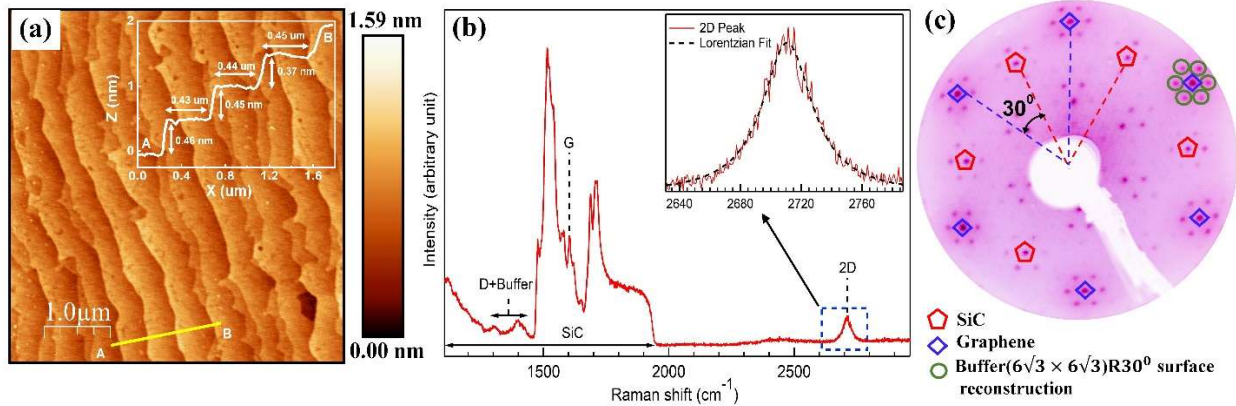

*Figure 8:* EG grown on Si-face of 4H-SiC in Open configuration: (a) A 5×5 μm AFM topography image of the EG/4H-SiC(0001). The height profile along the line (AB, yellow color) has been shown in the inset. (b) Raman spectrum of EG/4H-SiC(0001). A magnified view of the 2D peak region (marked by a dashed blue rectangle) is shown in the inset, along with a single Lorentzian fitting (dashed line). All the characteristic Raman features of graphene/4H-SiC are also marked. (c) Inverted LEED pattern of monolayer graphene/4H-SiC(0001) recorded at 100 eV electron beam

*energy. Diffraction spots corresponding to SiC (pentagons, red), a reconstructed buffer layer (open circles, green), and graphene (diamonds, blue) are also marked.*

wide range (1300 – 1950 cm$^{-1}$) area also marked. The line shape and full width at half maximum (FWHM) of the 2D peak provide crucial information on the layer thickness of graphene. The Inset of Fig. 8(b) shows the close-up view of the 2D peak, and it can be satisfactorily fitted by a single Lorentzian (dashed line, black). FWHM of 2D peak (40 cm$^{-1}$) and its peak position are in good agreement with expected values for monolayer EG.[18,26,59] LEED measurements were performed in multi-probe surface analysis UHV system from Omicron Nano Technology, and EG sample was annealed at 650°C (2 hours) to achieve a clean surface. The LEED pattern (Fig. 8(c)) of epitaxial monolayer graphene grown in open configuration shows characteristic diffraction spots of graphene lattice (diamonds, blue), SiC (pentagons, red), and (6√3×6√3) R30° surface reconstructed interface buffer layer (circles, green). Our LEED pattern is similar to the expected LEED pattern of monolayer EG,[18,19] and it confirms long-range surface periodicity and high crystalline quality of the grown graphene layer. The graphene lattice is expectedly rotated by 30° with respect to SiC.[18,19]

**B. Epitaxial graphene growth in CCS configuration:**

We utilized the CCS method within the GrapE system to facilitate a more precise control over Si escape rates, enabling a more regulated growth of epitaxial graphene (EG) on 4H-SiC. SiC ships were placed inside a specially designed graphite crucible (Fig. 3) to conduct EG growth in the CCS configuration. Crucially, the inner graphite walls of the CCS crucible needed passivation with Si before initiating actual EG growth runs. For this purpose, a few dummy SiC chips were heated in the CCS crucible before the final growth runs. The growth conditions, encompassing temperature, pressure, growth duration, and crucible geometry, were carefully tailored to attain the desired EG attributes such as layer thickness, crystalline quality, and uniformity. The growth process adhered to an optimized temperature/pressure profile illustrated in Fig. S6 (supplementary material). Figure 9(a) displays the AFM topographic image of the EG grown in CCS configuration, showcasing a relatively regular step-terrace-like morphology, an improvement compared to EG grown in the open configuration (Fig. 8(a)). The height profile along line AB (inset of Fig. 9(a)) highlights increased step bunching and reveals terrace widths varying between 3 - 6 μm, with step heights ranging from 8 - 10 nm, as evident from the height profile along the AB line (inset of Fig. 9(a)). Notably, no observable pit-like morphological defects indicate enhanced morphology in the CCS growth case compared to growth in the open configuration. The average *rms* surface roughness within individual EG terraces measures approximately ~0.035 nm. Figure 9(b) depicts the Raman spectrum, and characteristic G and 2D peaks are located at ~1606 and ~2726 cm$^{-1}$, respectively. Overlapping broad features with weak intensity, associated with defects and the interface buffer layer, are also identified. The tiny D peak indicates the high structural quality of the grown EG. The inset of Fig. 9(b) presents a close-up view of the 2D peak, satisfactorily fitted by a single Lorentzian (dashed line, black). The FWHM of the 2D peak (38 cm$^{-1}$) and its peak position align well with expected values for monolayer EG grown on SiC.[18,25,59] Figure 9(c) showcases the LEED pattern of monolayer EG produced in the CCS configuration, which closely resembles the LEED pattern of monolayer EG grown in the open configuration.

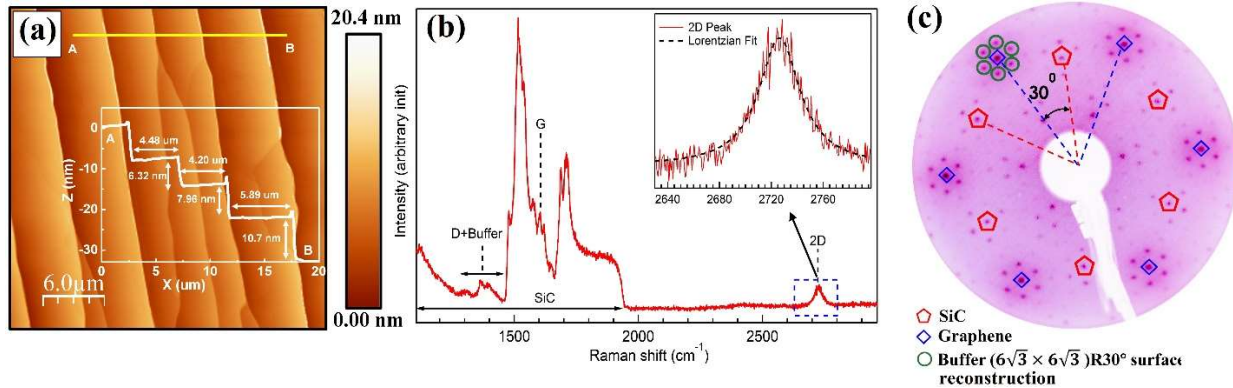

*Figure 9: EG grown on Si-face of 4H-SiC in confinement controlled sublimation (CCS) configuration: (a) A 5×5 μm AFM topography image of the EG/4H-SiC(0001). The height profile along the line (AB, yellow color) has been shown in the inset. (b) Raman spectrum of EG/4H-SiC(0001). A magnified view of the 2D peak region (marked by a dashed blue rectangle) is shown in the inset, along with a single Lorentzian fitting (dashed line). All the characteristic Raman features of graphene/4H-SiC are also marked. (c) Inverted LEED pattern of monolayer graphene/4H-SiC(0001) recorded at 100 eV electron beam energy. Diffraction spots corresponding to SiC (pentagons, red), a reconstructed buffer layer (open circles, green), and graphene (diamonds, blue) are also marked.*

**C. Epitaxial graphene growth in polymer-assisted CCS (PACCS) configuration:**

Highly uniform monolayer EG on SiC can be achieved at elevated growth temperatures. Typically, the received SiC surface can be prepared to exhibit step heights of ≤1 unit cell (u.c.). However, during EG sublimation growth at temperatures exceeding 1500°C, the step heights tend to increase (ranging between 5 and 15 nm) due to significant step bunching. EG growth primarily starts at step edges, and high step heights promote the inclusion of bilayers, leading to uneven monolayer coverage over a large area. Regions with elevated step heights manifest higher electrical resistance, causing greater electrical resistance anisotropy, which can degrade the performance of electronic devices based on EG. For applications like EG-based QHRS, maintaining shallow step heights (≤ 1 u.c. of SiC) is crucial, as elevated step heights hinder the performance of EG-based QHRS.[26,36] Therefore, achieving shallow step heights is essential for consistent and homogeneous monolayer EG, making it suitable for high-performance electronic devices, especially for QHRS. Polymer-assisted sublimation growth (PASG), introduced by Kruskopf *et al.* has emerged as one of the most effective methods to achieve monolayer EG on SiC with shallow step heights.[26] They coated the SiC chip with a photoresist polymer to provide an additional carbon source, resulting in the formation of a uniform buffer layer. Subsequent graphitization in an open configuration led to large-area coverage with monolayer EG exhibiting shallow step heights of ≤ 1 nm.[26] Monolayer EG with such low step heights, produced by PASG, demonstrated its applicability in metrology-grade QHRS at 4.2 K.[26, 36]

We employed the aforementioned growth philosophy in a similar approach but skipped the additional step involving polymer deposition/coating. We recall that our diced SiC chips are coated with AZ1518 photoresist. Instead of removing the polymer coating by ultrasonication in boiled acetone, we utilized the polymer already coated on the Si-face during dicing as a seed carbon source to control step bunching during EG growth. Additionally, the growth was performed in a CCS configuration for better control over the growth rate. We pre-cleaned polymer-coated diced SiC chips by only mild ultrasonic cleaning in isopropanol for 3-4 minutes and loaded these in CCS crucible to perform polymer-assisted CCS (PACCS) growth by following the same temperature/pressure growth recipe as used for open and CCS configuration growth (Fig. S6). Figure 10(a) displays the AFM surface topography image of EG grown on 4H-SiC(0001) using the PACCS method. It exhibits a regular step-terrace morphology, with terrace widths between 0.5 and 0.6 μm and shallow step heights ranging from 0.6 to 0.8 nm, as illustrated by the height profile given in the inset. The average *rms* surface roughness within the terraces is approximately ~0.02 nm, indicating atomically flat terraces. Our PACCS method achieves EG with similar shallow step heights to those reported for the previously used PASG method.[26] In Fig. 10(b), the Raman spectrum of EG grown using the PACCS method is depicted, highlighting the characteristic G and 2D peaks positioned at ~1610 and ~2752 cm$^{-1}$, respectively. The presence of a minimal D peak underscores the high structural quality of the produced EG. The inset of Fig. 10(b) provides a close- up view of the 2D peak, adequately fitted by a single Lorentzian (shown as a dashed black line). The FWHM of the 2D peak (~39 cm$^{-1}$) confirms the monolayer nature of EG grown via the PACCS method.[18,25,26,59] Additionally, we acquired a high-quality LEED pattern (Fig. S7, supplementary material) for the monolayer EG grown through the PACCS method. It closely mirrors the LEED pattern observed in monolayer EG grown via the open/CCS configuration, reinforcing the structural quality of the EG surface produced using the PACCS method.

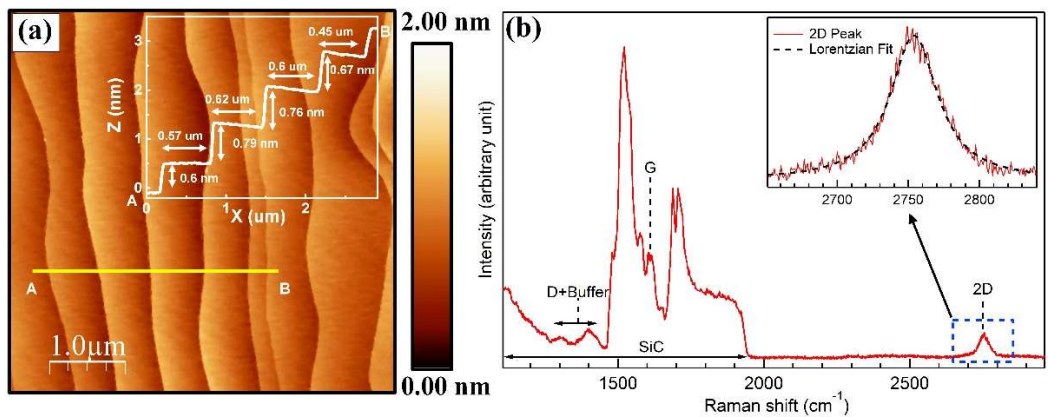

*Figure 10: EG grown on Si-face of 4H-SiC in polymer-assisted CCS (PACCS) configuration: (a) A 5×5 µm AFM topography image graphene/4H-SiC(0001). The height profile along the line (AB, yellow color) has been shown in the inset. (b) Raman spectrum of graphene/4H-SiC(0001). A magnified view of 2D peak region (marked by a dashed blue rectangle) is shown in the inset, along with a single Lorentzian fitting (dashed line). All the characteristic Raman features of graphene/4H-SiC are also marked.*

**D. Turbostratic graphene growth using RTA:**

Considering the importance of RTA in synthesizing turbostratic EG on C-face SiC, we explored the capabilities of the GrapE furnace in achieving high annealing rates. We employed a graphite crucible without radiation shielding graphite felt to ensure rapid heating and cooling during induction heating. However, the direct placement of the growth crucible on the quartz dome of the reaction chamber might potentially compromise the integrity of the quartz dome, particularly during extended periods of high-temperature heating (>1600°C). We devised a straightforward solution to address this issue: placing the CCS graphite crucible on a C-shaped quartz support within a quartz boat positioned in the quartz dome. This setup prevented direct contact between the graphite crucible and the quartz dome. A schematic representation of this arrangement is provided in Fig. S8(a) (supplementary material), while the RTA performance of the GrapE with this crucible arrangement is illustrated in Fig. S9 (supplementary material). Our experiments demonstrated the capability GrapE furnace to reach heating rates up to 200°C/s, maintaining stable temperatures as high as 1800°C (Fig. S9, supplementary material), under both high vacuum (HV) and 1 atm argon gas pressure. We want to mention that we have utilized only approximately one-third of the available RF power to achieve the mentioned ramp rates and temperatures. Consequently, if necessary, we anticipate that the GrapE system could reach even higher RTA temperatures and ramp rates. We prepared two EG samples on the C-face of 4H-SiC under different growth conditions: one in HV (~10$^{-6}$ mbar) and the other at 1 atm pressure of Ar gas. For clarity, we will refer to the samples grown in HV and 1 atm Ar gas pressure as TEG$_{HV}$ and TEG$_{Ar}$, respectively. The growth profile used for RTA growth (200°C/s @ 1800°C, 5 min) of TEG$_{Ar}$ in an Ar gas environment is depicted in Fig. S8(b) (supplementary material). The exact temperature profile was applied for TEG$_{HV}$, except the growth occurred without any Ar gas. Figure 11(a) shows the AFM topographic image of TEG$_{Ar}$ where triangular/hexagonal/trapezoidal shape graphene domains with 0.5 – 2.0 μm widths are separated by a network of ridges/wrinkles. This AFM image is characteristic of multilayer turbostratic EG grown on C-face SiC ($000\bar{1}$) where graphene growth occurs in island-fashion contrary to layer-by-layer fashion EG growth observed on Si-face SiC.[60,61] Corresponding height profile along AB line (inset of Fig. 11(a)) ridges/wrinkles with varying heights (1 to 15 nm), linked to thermal expansion mismatch between grown EG and SiC and consequent release of compressive stress.[60,61] Overall *rms* surface roughness calculated for 10×10 μm AFM image (Fig. 11(a)) turns out to be ~1.7 nm, whereas average *rms* surface roughness calculated within individual domains is ~0.15 nm. AFM topographic image of TEG$_{HV}$ is shown in Fig. 11(b). A comparison between Fig. 11(a) and (b) highlights evident differences in topography between TEG$_{Ar}$ and TEG$_{HV}$. TEG$_{HV}$ shows a notably rougher topography, displaying individual graphene domains or islands as irregularly scattered structures resembling burnt petals, each with a height of approximately ~10 – 15 nm. Unlike TEG$_{Ar}$, where wrinkles or ridges separate graphene domains, TEG$_{HV}$ exhibits randomly distributed deep (~30 nm) trenches or pits, mainly appearing at the boundaries of individual graphene domains, as evident from the height profile along AB line given in the inset of Fig. 11(b). The observed morphology of TEG$_{HV}$ aligns with an earlier report on turbostratic EG growth on the C-face of SiC.[47] The *rms* surface roughness calculated from a 10×10 μm AFM image (Fig. 11(b)) of TEG$_{HV}$ is ~7.8 nm, while the average *rms* surface roughness within individual domains measures around ~1.5 nm. The notably increased surface roughness of TEG$_{HV}$ can be attributed to the significantly augmented surface roughness or corrugation of the SiC surface at the initiation of RTA growth in a HV environment, where Si escape rates are anticipated to be much higher compared to growth performed in a 1 atm Ar gas backpressure.[47] The inverted low energy electron diffraction (LEED) patterns captured at a 70 eV beam energy for samples grown in atmospheric pressure of Ar gas (TEG$_{Ar}$) and in HV environments (TEG$_{HV}$) are depicted in Fig. 12. The LEED pattern of TEG$_{Ar}$ displays a bimodal six-fold symmetry pattern of diffuse arc sets (Fig. 12(a)). These elongated graphene arcs indicate arbitrary azimuthal rotations between individual graphene sheets of multilayer turbostratic EG grown on C-face SiC.[19,37] The smaller arc, marked by a blue dashed elliptical marker, exhibits a smaller angular distribution, positioned at an azimuthal angle Ø=0° relative to the SiC diffraction spot (red squares). In contrast, the larger diffused arc is rotated at an angle of 30° relative to the SiC spots. LEED is a surface-sensitive technique, and substrate LEED spots are usually extinguished after the growth of a few adlayers. However, in the case of island growth, it is not uncommon to observe substrate LEED spots even after multilayer growth. The presence of SiC LEED spots for TEG$_{Ar}$ confirms the previously reported island-like growth of this turbostratic EG on the C-face of SiC.[60]

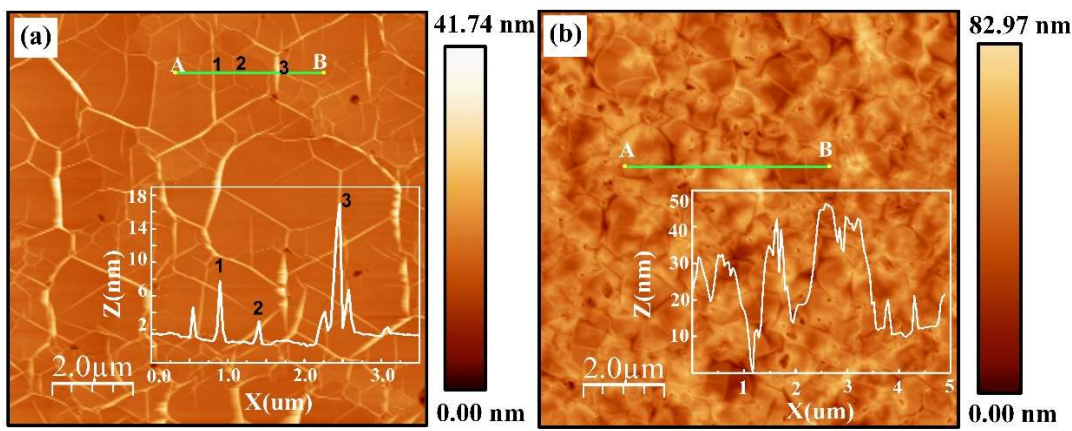

*Figure 11*: Turbostratic EG grown on C-face of 4H-SiC using RTA with 200 $^0$C/s heating rate. (a) and (b) show 10×10 μm AFM topography image of the graphene/4H-SiC ($000\bar{1}$) grown in 1 Atm Ar gas (TEG$_{Ar}$) and HV (TEG$_{HV}$), respectively. Respective height profiles for TEG$_{Ar}$ and TEG$_{HV}$ along the line (AB, green color) are shown as insets.

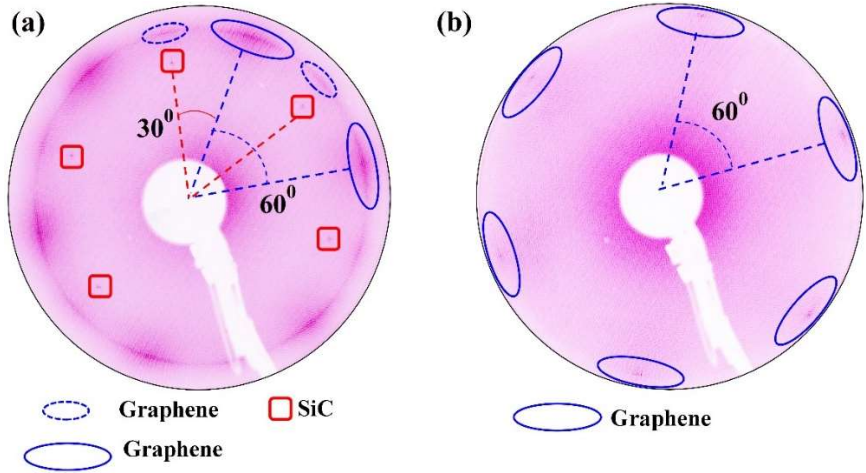

*Figure: 12:* Inverted LEED diffraction pattern of turbostratic EG grown on C—face 4H-SiC using RTA. LEED patterns were acquired at 70 eV beam energy. (a) LEED pattern of the sample (TEG$_{Ar}$) grown at 1atm Ar gas atmosphere, showing set of bimodal (small arcs with blue dotted elliptical marker and large diffused arc with blue solid elliptical marker) diffused arcs with sixfold symmetry along with SiC spots (red, solid square box). Small arcs and large diffused arcs are rotated by $\phi = 0°$ and 30° relative to SiC spots, respectively. (b) LEED pattern of the sample (TEG$_{HV}$) grown in a high vacuum atmosphere, depicting a set of monomodal diffused spots with six-fold symmetry (blue solid elliptical marker). No traces of SiC spots have been observed for TEG$_{HV}$.

Previous studies suggest that graphene islands on C-face SiC should grow to more than five layers thick before forming the first continuous graphene film covering the entire SiC substrate.[60] Significant differences between the LEED patterns of samples grown in high vacuum and atmospheric Ar gas pressure are observed from the comparison of Fig. 12(a) and (b). The LEED pattern of TEG$_{HV}$ exhibits a single set of diffused arcs with six-fold symmetry. The absence of SiC-related LEED spots for TEG$_{HV}$ indicates an increased thickness of graphene sheets due to higher growth rates in high vacuum. However, the higher thickness and significantly increased surface roughness caused by SiC surface roughening at high vacuum growth conditions lead to degradation of the LEED pattern for TEG$_{HV}$ compared to TEG$_{HV}$.

Figure 13(a) presents the Raman spectrum of TEG$_{HV}$, highlighting characteristic graphene features (D, G, and 2D). The absence of SiC-associated features confirms the multilayer nature of the grown EG. The D/G intensity ratio, ~0.06, indicates high structural quality with low defects. Despite being multilayered, the sharp 2D peak can be fitted with a single Lorentzian component (inset of Fig. 13(a)) exhibiting peak position and FWHM of 2701 cm$^{-1}$ and 44 cm$^{-1}$, respectively. Typically, in Bernal stacked monolayer EG, the 2D peak can be fitted with a single Lorentzian having FWHM ≤ 45 cm$^{-1}$. However, thicker Bernal stacked EG shows higher FWHM values, requiring multiple Lorentzian components for modeling the 2D peak line shape. In the case of each layer of a multilayer turbostratic EG with non-Bernal stacking, it behaves akin to a monolayer EG due to rotational decoupling. The 2D peak position of TEG$_{HV}$ indicates its turbostratic nature, and the FWHM value aligns closely with that expected for monolayer EG.[47,60] Conversely, the Raman spectrum of TEG$_{Ar}$ (Fig. 13(b)) exhibits SiC-related features, indicating lesser thickness compared to TEG$_{HV}$. The 2D peak (inset of Fig. 13(b)) cannot be satisfactorily fitted with a single Lorentzian, pointing to mixed stacking with contributions from rotationally disordered non-Bernal and ordered Bernal stackings. Three Lorentzian components (P1, P2, and P3 centered at ~2700, 2725, and 2678 cm$^{-1}$) are required to adequately fit the 2D peak (Fig. 13(c)), confirming the mixed stackings in TEG$_{Ar}$. The ratio of turbostratic stacking, calculated from the integrated areas of fitted Lorentzian components, stands at ~71%.[62,63]

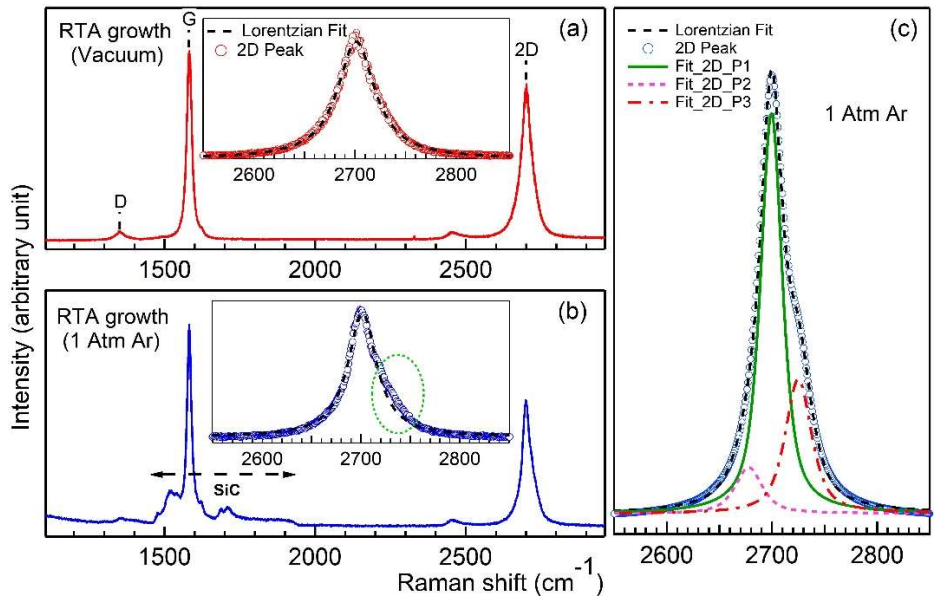

*Figure 13: Raman spectra for RTA grown turbostratic EG on C-face of 4H-SiC: (a) Raman spectrum of TEG$_{HV}$ grown in HV environment. The inset shows a close-up view of the 2D peak and the single Lorentzian fitting. (b) Raman spectrum of TEG$_{Ar}$ grown at 1 atm Ar gas pressure. The inset shows a magnified view of the 2D peak and a single Lorentzian fitting, and the dashed oval shows a less satisfactory fit region. (c) 2D peak region of TEG$_{Ar}$ along with three Lorentzian fitting components. Ticks identify all the characteristic Raman features of EG, and region having overlapping SiC features is marked by a dashed line with arrows.*

Notably, the change in growth ambient from HV to atmospheric pressure of Ar gas during RTA leads to a shift from pure turbostratic stacking to a mixed stacking order. The stacking order of multilayer graphene significantly impacts its electronic and optical properties. Thus, further investigations into tuning the RTA growth of turbostratic EG at various combinations of Ar gas pressure and heating rates may allow precise control over the mixing of non-Bernal/Bernal stacking orders, offering avenues for future exploration. The GrapE system, with its RTA capabilities in different growth settings, is perfectly suited for conducting such studies.

## IV. Conclusion:

We have presented *grap*hene *e*pitaxy (GrapE) system, an RF induction heating-based furnace designed for epitaxial graphene (EG) growth on SiC. GrapE fulfills essential criteria for EG growth, offering controlled growth environments from high vacuum to atmospheric pressures, stable temperatures up to ~2000°C, and controlled heating rates from 1°C/s to 200°C/s. We cover its design, construction, thermal characteristics, and EG growth performance extensively. The GrapE furnace demonstrates its versatility by utilizing diverse EG growth methods. The automated furnace ensures high-quality EG production on SiC. Furthermore, we introduce a modified CCS method—polymer-assisted CCS—for EG growth with shallow step heights on SiC (0001). Moreover, we employ rapid thermal annealing (RTA) with a high annealing rate of 200°C/s to generate turbostratic EG on SiC ($000\bar{1}$) and investigate the influence of varying growth environments on its stacking order. The produced EG quality is evaluated using atomic force microscopy, Raman spectroscopy, and low-energy electron diffraction.

## ACKNOWLEDGEMENTS:

This work has been supported by OLP projects (170232, 180232, 210232, and 230232) and the mission mode project (HCP-55) funded by CSIR. SM acknowledges the financial support from DST-INSPIRE fellowship. The authors thank to Dr. P. K. Siwach, Ms. Jyoti, and Mr. Aakash for their assistance in setting up the GrapE system. The authors extend their sincere gratitude to Mr. Suresh, Mr. Ramdhan, Mr. Manojit, Mr. Rajan, Mr. Pawan, Mr. V. K. Gupta, and Dr. S. K. Jaiswal for their valuable help and support in fabricating various parts of the GrapE furnace. The authors also thank Dr. S. P. Singh, Dr. V. V. Agarwal, and Dr. V. K. Jaiswal for facilitating AFM and Raman spectroscopy measurements. Mr. Surjit Sardar is thanked for his help in automation of GrapE furnace. The authors are also grateful for the consistent encouragement and support received from the Director of CSIR-NPL and Dr. H. K. Singh.

## AUTHOR DECLARATIONS

### Conflict of Interest

The authors have no conflicts to disclose.

### Authors contribution:

**S. Mondal:** Conceptualization (equal); Data curation (equal); Formal analysis (equal); Investigation (equal); Methodology (equal); Project administration (supporting); Software (equal); Validation (equal); Visualization (equal); Writing – original draft (equal); Writing – review & editing (equal). **U. J. Jayalakshmi:** Data curation (supporting), Resources (supporting). **S. Singh:** Data curation (supporting), Resources (supporting). **R. K. Mukherjee:**, Resources (supporting). **A.K. Shukla:** Conceptualization (equal); Data curation (equal); Formal analysis (equal); Investigation (equal); Methodology (equal); Project administration (lead); Software (equal); Validation (equal); Visualization (equal); Writing – original draft (equal); Writing – review & editing (equal)

## DATA AVAILABILITY:

The data that support the findings of this study are available from the corresponding author upon reasonable request

## REFERENCES:

[1] K.S. Novoselov, A.K. Geim, S. V. Morozov, D. Jiang, Y. Zhang, S. V. Dubonos, I.V. Grigorieva, and A.A. Firsov, Science **306**, 666 (2004).

[2] A.K. Geim, Science **324**, 1530 (2009).

[3] A.H. Castro Neto, F. Guinea, N.M.R. Peres, K.S. Novoselov, and A.K. Geim, Rev. Mod. Phys. **81**, 109 (2009).

[4] Y. Zhang, Y.W. Tan, H.L. Stormer, and P. Kim, Nature **438**, 201 (2005).

[5] A.C. Ferrari *et al.*, Nanoscale **7**, 4598 (2015).

[6] K.S. Novoselov, V.I. Fal'Ko, L. Colombo, P.R. Gellert, M.G. Schwab, and K. Kim, Nature **490**, 192 (2012).

[7] G.R. Yazdi, T. Iakimov, and R. Yakimova, Crystals **6**, 53 (2016).

[8] Y.H. Wu, T. Yu, and Z.X. Shen, J. Appl. Phys. **108**, 071301 (2010).

[9] K.S. Novoselov, A.K. Geim, S. V. Morozov, D. Jiang, M.I. Katsnelson, I. V. Grigorieva, S. V. Dubonos, and A.A. Firsov, Nature **438**, 197 (2005).

[10] K.S. Novoselov, Z. Jiang, Y. Zhang, S. V Morozov, H.L. Stormer, U. Zeitler, J.C. Maan, G.S. Boebinger, P. Kim, and A.K. Geim, Science **315**, 2007 (2007).

[11] F. Lafont, R. Ribeiro-Palau, Z. Han, A. Cresti, A. Delvallée, A.W. Cummings, S. Roche, V. Bouchiat, S. Ducourtieux, F. Schopfer, and W. Poirier, Phys. Rev. B **90**, 115422 (2014).

[12] I. Forbeaux, J. Themlin, and J. Debever, Phys. Rev. B **58**, 16396 (1998).

[13] C. Berger, Z. Song, T. Li, X. Li, A.Y. Ogbazghi, R. Feng, Z. Dai, N. Alexei, M.E.H. Conrad, P.N. First, and W.A. De Heer, J. Phys. Chem. B **108**, 19912 (2004).

[14] A. Bostwick, K. V. Emtsev, K. Horn, E. Huwald, L. Ley, J.L. McChesney, T. Ohta, J. Riley, E. Rotenberg, F. Speck, and T. Seyller, Adv. Solid State Phys. **47**, 159 (2008).

[15] L.O. Nyakiti, V.D. Wheeler, N.Y. Garces, R.L. Myers-Ward, C.R. Eddy, and D.K. Gaskill, MRS Bull. **37**, 1149 (2012).

[16] S.Y. Zhou, G.H. Gweon, A. V. Fedorov, P.N. First, W.A. De Heer, D.H. Lee, F. Guinea, A.H. Castro Neto, and A. Lanzara, Nat. Mater. **6**, 770 (2007).

[17] T. Seyller, A. Bostwick, K. V. Emtsev, K. Horn, L. Ley, J.L. McChesney, T. Ohta, J.D. Riley, E. Rotenberg, and F. Speck, Phys. Status Solidi Basic Res. **245**, 1436 (2008).

[18] K. V. Emtsev, A. Bostwick, K. Horn, J. Jobst, G.L. Kellogg, L. Ley, J.L. McChesney, T. Ohta, S.A. Reshanov, J. Röhrl, E. Rotenberg, A.K. Schmid, D. Waldmann, H.B. Weber, and T. Seyller, Nat. Mater. **8**, 203 (2009).

[19] W.A. De Heer, C. Berger, M. Ruan, M. Sprinkle, X. Li, Y. Hu, B. Zhang, J. Hankinson, and E. Conrad, Proc. Natl. Acad. Sci. **108**, 16900 (2011).

[20] S.K. Lilov, Mater. Sci. Eng. B **21**, 65 (1993).

[21] G. Honstein, F. Baillet, and C. Chatillon, J. Eur. Ceram. Soc. **32**, 4407 (2012).

[22] R. Yakimova, G.R. Yazdi, T. Iakimov, J. Eriksson, and V. Darakchieva, ECS Trans. **53(1)**, 9 (2013).

[23] C. Virojanadara, M. Syväjarvi, R. Yakimova, L.I. Johansson, A.A. Zakharov, and T. Balasubramanian, Phys. Rev. B **78**, 245403 (2008).

[24] M. Ostler, F. Speck, M. Gick, and T. Seyller, Phys. Status Solidi Basic Res. **247**, 2924 (2010).

[25] M.A. Real, E.A. Lass, F.H. Liu, T. Shen, G.R. Jones, J.A. Soons, D.B. Newell, A. V. Davydov, and R.E. Elmquist, IEEE Trans. Instrum. Meas. **62**, 1454 (2013).

[26] Y. Yang, G. Cheng, P. Mende, I.G. Calizo, R.M. Feenstra, C. Chuang, C. Liu, C. Liu, G.R. Jones, A.R. Hight, and R.E. Elmquist, Carbon **115**, 229 (2017).

[27] M. Kruskopf *et al.*, 2D Mater. **3**, 041002 (2016).

[28] N. Zebardastan, J. Bradford, J. Lipton-Duffin, J. MacLeod, K. Ostrikov, M. Tomellini, and N. Motta, Nanotechnology **34**, (2023).

[29] M. Kruskopf, Epitaxial Graphene on SiC for Quantum Resistance Metrology, Physikalisch-Technische Bundesanstalt (PTB), *PhD dissertation*, Braunschiweig and Berlin, 2017.

[30] J. H. Hankinson, Spin Dependent Current Injection Into Epitaxial Graphene Nanoribbons, *PhD dissertation*, Georgia Institute of Technology, 2015.

[31] D. Deniz, Fabrication of Arrays of Ballistic Epitaxial Graphene Nanoribbons, *PhD dissertation*, Georgia Institute of Technology, 2018.

[32] J. Turmaud, Variable Range Hopping Conduction in the Epitaxial Graphene Buffer Layer on SiC (0001 ), *PhD dissertation*, Georgia Institute of Technology, 2018.

[33] J. Kunc, M. Rejhon, E. Belas, V. Dědič, P. Moravec, and J. Franc, Phys. Rev. Appl. **8**, 044011 (2017).

[34] E. Pallecchi, F. Lafont, V. Cavaliere, F. Schopfer, D. Mailly, W. Poirier, and A. Ouerghi, Sci. Rep. **4**, 4558 (2014).

[35] A. Lartsev, T. Yager, T. Bergsten, A. Tzalenchuk, T.J.B.M. Janssen, R. Yakimova, S. Lara-Avila, and S. Kubatkin, Appl. Phys. Lett. **105**, 063106 (2014).

[36] M. Kruskopf and R.E. Elmquist, Metrologia **55**, R27 (2018).

[37] J. Hass, F. Varchon, J.E. Millán-Otoya, M. Sprinkle, N. Sharma, W.A. De Heer, C. Berger, P.N. First, L. Magaud, and E.H. Conrad, Phys. Rev. Lett. **100**, 3 (2008).

[38] C. Wei, R. Negishi, Y. Ogawa, M. Akabori, Y. Taniyasu, and Y. Kobayashi, Jpn. J. Appl. Phys. **58**, SIIB04 (2019).

[39] R. Negishi, C. Wei, Y. Yao, Y. Ogawa, M. Akabori, Y. Kanai, K. Matsumoto, Y. Taniyasu, and Y. Kobayashi, Phys. Status Solidi Basic Res. **257**, 1900437 (2020).

[40] J.A. Garlow, L.K. Barrett, L. Wu, K. Kisslinger, Y. Zhu, and J.F. Pulecio, Sci. Rep. **6**, 19804 (2016).

[41] K.F. Mak, M.Y. Sfeir, J.A. Misewich, and T.F. Heinza, Proc. Natl. Acad. Sci. **107**, 14999 (2010).

[42] J.J. Yoo, K. Balakrishnan, J. Huang, V. Meunier, B.G. Sumpter, A. Srivastava, M. Conway, A.L. Mohana Reddy, J. Yu, R. Vajtai, and P.M. Ajayan, Nano Lett. **11**, 1423 (2011).

[43] N. Gupta, U. Mogera, and G.U. Kulkarni, Mater. Res. Bull. **152**, 111841 (2022).

[44] P. Kokmat, P. Surinlert, and A. Ruammaitree, ACS Omega **8**, 4010 (2023).

[45] Z. Liu, Q. Xu, C. Zhang, Q. Sun, C. Wang, M. Dong, Z. Wang, H. Ohmori, M. Kosinova, T. Goto, R. Tu, and S. Zhang, Appl. Surf. Sci. **514**, 145938 (2020).

[46] D. Han, X. Wang, Y. Zhao, Y. Chen, M. Tang, and Z. Zhao, Carbon **124**, 105 (2017).

[47] T. Hu, H. Bao, S. Liu, X. Liu, D. Ma, F. Ma, and K. Xu, Carbon **120**, 219 (2017).

[48] J. Prekodravac, Z. Marković, S. Jovanović, M. Budimir, D. Peruško, I. Holclajtner-Antunović, V. Pavlović, Z. Syrgiannis, A. Bonasera, and B. Todorović-Marković, Synth. Met. **209**, 461 (2015).

[49] R.S. Rajaura, I. Singhal, K.N. Sharma, and S. Srivastava, Rev. Sci. Instrum. **90**, 123903 (2019).

[50] C. Zhang, J. Zhang, K. Lin, and Y. Huang, Rev. Sci. Instrum. **88**, 053907 (2017).

[51] F. Pang, Rev. Sci. Instrum. **89**, 086104 (2018).

[52] T. Aritsuki, T. Nakashima, K. Kobayashi, Y. Ohno, and M. Nagase, Jpn. J. Appl. Phys. **55**, 06GF03 (2016).

[53] R.R. Wilson, Rev. Sci. Instrum. **12**, 91 (1941).

[54] V. Stanishev, N. Armakavicius, C. Bouhafs, C. Coletti, P. Kühne, I.G. Ivanov, A.A. Zakharov, R. Yakimova, and V. Darakchieva, Appl. Sci. **11**, 1891 (2021).

[55] Ö. Danielsson, U. Forsberg, A. Henry, and E. Janzén, J. Cryst. Growth **235**, 352 (2002).

[56] T. Chiba, S. Yamada, and E. Otsuki, J. Magn. Soc. Japan **22**, S1_301 (1998).

[57] A.K. Shukla, S. Banik, R.S. Dhaka, C. Biswas, S.R. Barman, and H. Haak, Rev. Sci. Instrum. **75**, 4467 (2004).

[58] Zenmansk and M.W. Y, *Heat and Thermodynamics*, 5th ed. (Mc Graw Hill, Kogakusha,Tokyo, 1968).

[59] M. Kruskopf, K. Pierz, S. Wundrack, R. Stosch, T. Dziomba, C.C. Kalmbach, A. Müller, J. Baringhaus, C. Tegenkamp, F.J. Ahlers, and H.W. Schumacher, J. Phys. Condens. Matter **27**, (2015).

[60] C.E. Giusca, S.J. Spencer, A.G. Shard, R. Yakimova, and O. Kazakova, Carbon **69**, 221 (2014).

[61] N. Camara, J.R. Huntzinger, G. Rius, A. Tiberj, N. Mestres, F. Pérez-Murano, P. Godignon, and J. Camassel, Phys. Rev. B **80**, 125410 (2009).

[62] L.G. Cançado, K. Takai, T. Enoki, M. Endo, Y.A. Kim, H. Mizusaki, N.L. Speziali, A. Jorio, and M.A. Pimenta, Carbon **46**, 272 (2008).

[63] R. Negishi, K. Yamamoto, H. Tanaka, S.A. Mojtahedzadeh, N. Mori, and Y. Kobayashi, Sci. Rep. **11**, 10206 (2021).